# Attojoule superconducting thermal logic and memories

Hui Wang[1], Niels Noordzij[2], Stephan Steinhauer[3], Thomas Descamps[3], Eitan Oksenberg[2], Val Zwiller[3] & Iman Esmaeil Zadeh[1,2]

**Due to stringent thermal budgets in cryogenic technologies such as superconducting quantum computers and sensors, minimizing the energy dissipation and power consumption of cryogenic electronic components is pivotal for large-scale devices. However, electronic building blocks that simultaneously offer low energy consumption, fast switching, low error rates, a small footprint and simple fabrication remain elusive. In this work, we demonstrate a superconducting switch with attojoule switching energy, high speed (pico-second rise/fall times), and high integration density (on the order of $10^{-2}$ μm$^2$ per switch). The switch consists of a superconducting channel and a metal heater separated by an insulating silica layer, prepared using lift-off techniques. We experimentally demonstrate digital gate operations utilizing this technology, such as NOT, NAND, NOR, AND, and OR gates, with a few femtojoule energy consumption and ultralow bit error rates < $10^{-8}$. In addition, we build volatile memory elements with few femtojoule energy consumption per operation, a nanosecond operation speed, and a retention time over $10^5$ s. These superconducting switches open new possibilities for increasing the size and complexity of modern cryogenic technologies.**

[1] *Department of Imaging Physics, Delft University of Technology, 2628CN Delft, The Netherlands.*
[2] *Single Quantum B.V., 2628 CH Delft, The Netherlands.*
[3] *Department of Applied Physics, Royal Institute of Technology (KTH), SE-106 01 Stockholm, Sweden*

## 1. Introduction

Superconducting circuits have become pivotal to a wide range of computing and sensing applications such as superconducting qubits and quantum processors[1,2], infrared superconducting single-photon detectors[3], and superconducting neurons[4,5], because they can provide quantum state generation and manipulation with fast response and low dissipation. However, scaling up the current technology to a practical superconducting system is hampered by the limited cooling budget of the cryogenic working environment available to deal with the heat load caused by both signals and readout electronics.[6] Thus, developing low-power-consumption electronics becomes increasingly attractive as it can enhance the information processing capacity of superconducting quantum computers and sensors.

A great research effort has been invested in developing compatible cryogenic logic gates and memory cells, which are the building blocks of computing circuits, such as single-flux-quantum (SFQ) technology[7-11] or nanocryotrons (nTrons)[12,13]. However, the various technologies to realize these building blocks all present different combinations of challenges, such as external magnetic field sensitivity[14,15], need for additional timing circuits[16,17], large cell dimensions[18,19], destructive readout schemes[13], insufficient operation speeds[13,20], etc. As a result, there remains a demand for new technological concepts to realize both logic gates and memory cells for large-scale, complex cryogenic technologies.

In this article, we characterize the static and transient responses of a novel superconducting thermal switch actuated by a metal heater on top of a superconducting channel, as shown in Figure 1. We show that, under proper operation conditions, the required switching energy of our devices is no more than a few hundred attojoules. Thanks to the electrical insulation between the input and output channels, the superconducting thermal switch can be easily integrated with similar switches or other cryogenic devices. Using these compact nanostructures, we successfully implement digital logic operations (NOT, NOR, NAND, AND, and OR) and highlight the technology's potential for increased energy efficiency, fast operation speed, and large-scale and high-density integration. Our experimental results show that our logic devices can perform with an energy consumption of a few femtojoules, an operation speed of 100 MHz, and an excellent bit error rate (BER) of less than $10^{-8}$. While superconducting nanowire memory cells based on flux trapping have been demonstrated,[21,22] we present an ultracompact and fabrication-friendly approach to operate a single superconducting thermal switch as a volatile memory cell. This technology is based on the persistence of the resistive state in the superconducting channel due to excessive self-heating effect.[23] With its non-destructive readout scheme, fast operation speed (a few nanoseconds for Write or Reset operations, and sub-nanosecond for Read operations) and high reliability (< $10^{-6}$ error rate), the resulting superconducting memory cell shows great promise for scalable and energy-efficient memory arrays in cryogenic electrical systems.

## 2. Main

**Superconducting thermal switch**

The structure of the superconducting thermal switch is depicted in Figure 1a. A metal (Ti) heater is placed over a superconducting (NbTiN) channel of 8-10 nm thickness, with a thin insulating (SiO$_2$) layer in between. As illustrated in Figure 1c, the local heat profiles are determined by joule heat generation in the metal heater and phonon transport through the SiO$_2$ layer. As the local temperature exceeds the critical temperature $T_c$, the superconducting state of the NbTiN channel is disrupted, resulting in a resistive section ("Off state" in Figure 1c). The resistance in the NbTiN channel is determined by the area of the resistive region, which is related to both the current through the heater, $I_h$, and the bias current of the NbTiN channel, $I_b$.

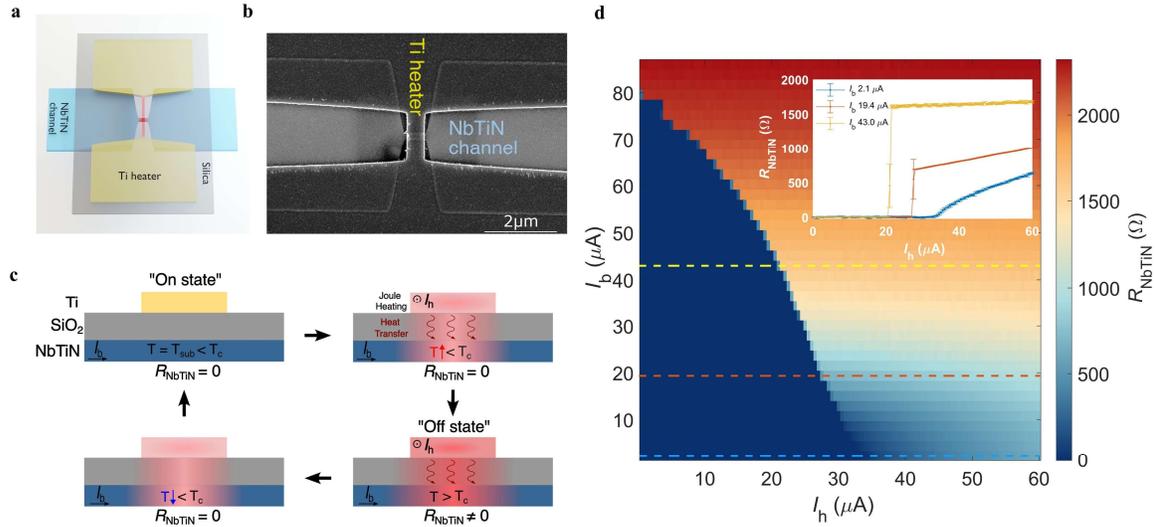

Figure 1: Characterization of the superconducting thermal switch. **a,** Schematic picture and **b,** Scanning Electron Microscopy image of the superconducting thermal switch consisting of a NbTiN channel covered by a layer of silica and a Ti heater. **c,** The device switches between the superconducting ("On") state and the resistive ("Off") state due to heat transfer from the Ti via the silica to the NbTiN channel. **d,** Resistance of the NbTiN channel ($R_{NbTiN}$) as a function of the bias current ($I_b$) and the current through the metal heater ($I_h$).

Devices were fabricated as detailed in Methods. We first measured the response of the superconducting thermal switch in steady state. Direct currents $I_h$ and $I_b$ were sent through the Ti heater and the NbTiN channel, respectively, and the resistance of the NbTiN channel $R_{NbTiN}$ was measured (see Methods). Figure 1d displays the results for a device with a NbTiN constriction of 0.5 μm (width) × 0.1 μm (length) and a Ti heater of 0.1 μm (width) × 0.5 μm (length). The Figure shows that increasing the bias current reduces the local critical temperature, allowing for a lower switching current. Comparing the response curves at various bias currents shown in the inset, we can observe an abrupt change at the switching point between the superconducting and the resistive state when the bias current $I_b$ is high. This indicates that the resistive hot spot rapidly expands over the entire constriction due to the additional joule heat generation by the constant bias in the NbTiN channel.

In order to evaluate the switching behavior of this device, we measured the transient response. A schematic of the measurement circuit is displayed in Figure 2a. When the input pulse amplitude on the heater triggers a local hot spot, i.e. the superconducting thermal switch is in the off state, the resistance of the NbTiN channel $R_{\text{NbTiN}}$ becomes large enough to redirect the current to the output amplifier, producing a high voltage level at the output. The dynamic switching energy of the device is calculated as $E_h = \tau_{\text{in}} V_{\text{in}}^2 / R_h$, where $\tau_{\text{in}}$ and $V_{\text{in}}$ are the duration and amplitude of a single input pulse, respectively.

Figure 2b shows a typical example of the input and output waveforms. It should be noted that the time delay between the input and the output signals depends on the length of the associated cables and the differences between electronic components used in the experiment. Here, for easier comparison, the pulse traces are shifted to overlap. The superconducting thermal switch was biased at $I_b = 28.5$ µA and could be switched by an input pulse trace with a duration of $\tau_{\text{in}} \approx 4.2$ ns and a frequency of 10 MHz, corresponding to a dynamic switching energy of $E_h = 746.9$ aJ. The lowest dynamic switching energy measured was $E_h = 273.8$ aJ when the NbTiN channel was biased close to the critical current (see Supplementary Figure S3a). Such dynamic energy consumption is two to three orders of magnitude lower than that of the modern CMOS switches[24]. Considering the source-switch impedance mismatches and the connecting transmission lines, the estimated energy consumption presents a higher bound and may be much better (see Methods). Taking into account the energy dissipation due to the current flowing through the channel, the entire switch was estimated to consume 1.658 fJ every switching operation. Further improvement of the energy efficiency can be achieved by scaling down the device dimension to reduce the amount of energy required to break the superconductivity.

The rise and the fall time of the transient response in Figure 2b are 271.8 ps and 339.1 ps, respectively. The inductor time constant $L_k/R$ is not the main limiting factor due to the small NbTiN channel dimensions ($L_k$ is estimated to be few nH). Constrained by the temporal characteristics of the instrumentation (details in Methods), the highest switching frequency we experimentally applied was 200 MHz (see Supplementary information). It can be observed that the output pulse duration in Figure 2b does not fully match the input pulse duration. This could be due to the switch-on delay time and the thermal recovery time after the heater switches off. Previous studies of similar multilayer structures suggest that the switch-on delay could be decreased with a higher input power density or a higher bias current, which influences how fast the channel can be thermalized to the critical temperature.[20,25] The thermal recovery time is affected by the self-heating effect in the superconducting channel after deactivating the heater. Apart from engineering the material properties to obtain better thermal coupling between the channel and the substrate, a feasible way to alleviate it is to select a proper constriction geometry as well as an appropriate electrical bias level to reduce the self-heating power density.[26] A more thorough understanding of the working process requires an in-depth investigation of the material properties and electro-thermal dynamics of the three-layer system, including the phonon generation within the heater, the phonon diffusion from the metal heater to the superconducting film via the thin dielectric layer, the interaction between phonons and quasiparticles in the superconducting film, and phonon escape from the superconducting film to the substrate.[27]

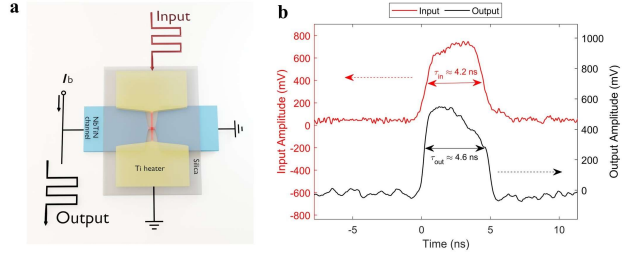

Figure 2: Transient response of the superconducting thermal switch. **a,** Schematic of the experimental circuit to test the transient response of the superconducting thermal switch. **b,** Voltage traces of a superconducting thermal switch with a clock input signal. The input signal is a 10 MHz square wave with a pulse duration $\tau_{in} \approx 4.2$ ns. The device is biased with $I_b = 28.5$ µA and the Ti heater has a resistance $R_h = 291.5$ Ω.

**Logic gates**

By controlling the resistance of the NbTiN channel, we designed a set of fundamental logic gates e.g. a NOT, NOR, NAND, AND, and OR gate. Figure 3 shows the schematics of the experimental circuits for NOT, NOR, and NAND gates, and the corresponding rescaled waveforms based on the logic LOW/HIGH state. More examples can be found in the Supplementary Figure S4.

In the NOT gate, a voltage source $V_{dc}$ is biased through the NbTiN channel of the superconducting thermal switch, in series with a 50 Ω load ($R_{Load}$ in Figure 3a). When the voltage across the metal heater is in the logic HIGH level, the NbTiN channel is switched to the non-superconducting state, introducing a nonzero resistance (usually ~kΩ) in the circuit. Therefore, the current through the load resistor is reduced when the superconducting thermal switch is off, leading to the logic LOW level at the output, as shown in Figure 3a. The bias voltage results in a power dissipation of $V_{dc}^2/(R_0 + R_{Load}) \approx 13.9$ nW in the on state. Since the normal resistance generated in the NbTiN channel is on the order of few hundred Ohm according to the static measurement, the energy dissipated in the channel and the bias circuit in the off state (approximated as $V_{dc}^2 \tau_A/(R_0 + R_{Load} + R_{NbTiN})$, where $\tau_A = 4.2$ ns is the pulse duration) is negligible in comparison with the heater input energy. Therefore, the total off-state energy dissipation is around 1.43 fJ.

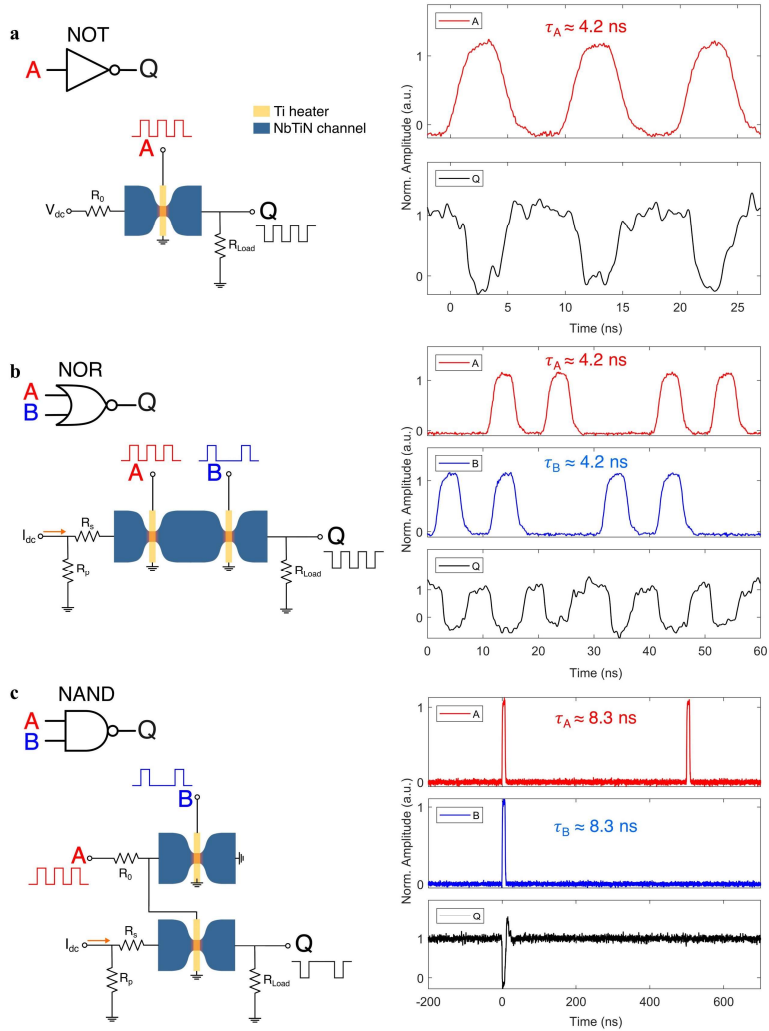

Figure 3: Construction of fundamental logic gates. **a-c,** Implementation of fundamental logic gates NOT (a), NOR (b), and NAND (c) using superconducting thermal switches. Figures on the left are the schematic diagrams. Figures on the right are the corresponding experimental waveforms of the inputs (A, B) and output (Q), rescaled according to the logic HIGH and LOW levels. The resistive elements shown in the figure are $R_0 = 22\ \Omega$, $R_{\text{Load}} = 50\ \Omega$, $R_p = 100\ \Omega$, and $R_s = 50\ \Omega$.

To implement a NOR gate, two superconducting thermal switches were connected in series (Figure 3b). Once either of the NbTiN channel switches to the resistive state, more current is redistributed to the parallel resistor $R_\text{p}$, leading to a logic LOW level across the output resistor $R_\text{Load}$. Since both switches have significantly larger normal-state resistances than the parallel resistor, there is only a minor difference between the low voltage levels with only one switch or both switches in resistive state. Compared with the NOT gate, a higher static power was dissipated due to the resistor bridge ($I_\text{dc}^2 [R_\text{p} \mathbin{/\mkern-5mu/} (R_\text{s} + R_\text{Load})] \approx 54.5$ nW) in the on state. The energy dissipation in the biasing and observation circuit is approximately 0.457 fJ, and the input heating pulse energies on the two heaters are 1.54 fJ and 1.53 fJ, respectively. Thus, the resulting energy dissipation of the whole NOR gate in the off state is around 3.53 fJ.

Analogously, one can assemble two superconducting thermal switches in parallel to build a NAND gate. This implementation requires an appropriate gate threshold or a delicate circuit design (see Supplementary Figure S5). Figure 3c represents another more straightforward way to construct a NAND gate, by connecting a NOT gate with an AND gate. Thanks to the insulation between the superconducting channel and the heater, the bias current and the heater current can be modulated simultaneously without considerable crosstalk. Therefore, the heater of a superconducting switch can be driven by the input signal B, while the bias current through the superconducting channel can be controlled by the other input signal A, forming an AND gate. The overshoot at the output, which can be observed in Figure 3c (as well as for the two-stage logic circuit in Figure 4b), can be attributed to the off-chip connection using coaxial cables between two stages in the initial experiments. During the switching-off time of 8.3 ns, the energy dissipated in the NAND gate is about 18.83 fJ. This number can be further decreased by using a voltage source for the NOT gate to avoid the static power consumption in the resistor bridge, integrating elements completely on-chip to increase the signal transfer efficiency, or shrinking the size of the superconducting thermal switch to achieve less energy dissipation.

One crucial merit of any logic gate is the bit error rate (BER), which is defined as the number of incorrect responses at the output divided by the total number of input events. We achieved a BER of $4.99\times10^{-9}$ for the NOR gate operation, as shown in Figure 4a. The two superconducting thermal switches were operated with input pulse energies of 1.40 fJ (input A) and 1.18 fJ (input B). The single fail event among over 200 million measurements was detected by comparing the time delay between the two consecutive output pulses with the supposed period of the output signal, which was 10 ns for Figure 4a.

The construction of NAND gates by combining NOT and AND gates suggests that the superconducting thermal switches can drive consecutive logic gates, which is important in building more complex superconducting logic circuits. To further explore this, we constructed the circuit shown in Figure 4b, where a NOT gate and a NOR gate are connected to form a simple logic circuit in such a way that one gate drives the input (heater) of the next stage. The determining factor to trigger the heater in the next stage is to generate a HIGH voltage level exceeding its switching threshold. Since the switch in the NOT gate is supposed to stay in the superconducting state at logic HIGH output, the HIGH voltage level is mainly constrained by the critical current of the NOT gate. Thus, the superconducting thermal switch used in the NOT gate was designed to have a wider NbTiN channel in order to provide enough switching energy to the next stage.

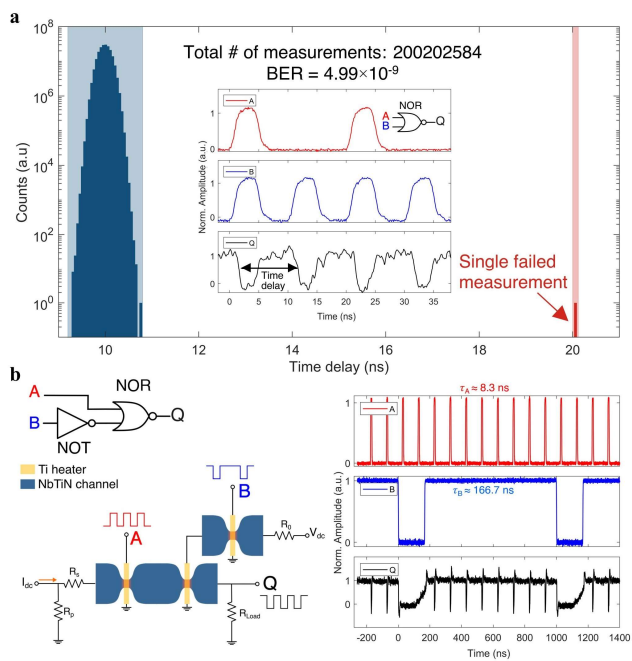

Figure 4: Bit error rate (BER) and driving ability of the logic gates. **a,** Correct (blue) and incorrect (red) operations of a NOR gate over $2 \times 10^8$ measurements as used to calculate the bit error rate (BER). The inset displays typical input (A: 50 MHz with a pulse duration of 4.2 ns, B: 100 MHz with a pulse duration of 4.2 ns) and output signals (Q) of the NOR gate. The horizontal coordinate represents the time delay between the falling edges of two consecutive pulses at the output, which was triggered at 50% of the output pulse amplitude. **b,** Electrical diagram (left) and experimental results (right) of a simple logic circuit composed of a NOT gate and a NOR gate.

**Memory cell**

Recently superconducting memory cells driven by both optical and electrical signals have been demonstrated using the hysteretic transition between the superconducting and the resistive state.[23] These devices consume ~600 μW power in the resistive state and operate at speeds < 50 kHz. In Figure 5, we present a novel approach to construct a volatile memory cell with the superconducting thermal switch, which allows a higher speed and lower energy consumption (see Supplementary Figure S6 for alternative implementations).

As depicted in Figure 5, the channel in a superconducting thermal switch can be intentionally latched into the non-superconducting state (state "1" in Figure 5a), if the electro-thermal effect in the superconducting channel and the heat dissipation into the substrate reach an equilibrium. By reducing the power dissipated from the channel or the heater to alleviate the electro-thermal effect, the device can be reset back to the superconducting state (state "0" in Figure 5a).

In the example shown in Figure 5b, the Write or Reset operations are realized by increasing the heater voltage $V_h$ or decreasing the bias voltage $V_b$. The "0" and "1" states of the device are distinguished based on the response at the output by sending a small pulse through the channel. When reading the resistive ("0") state, the output voltage level remains stable since $R_{NbTiN} \neq 0$ is dominant compared to the associated resistance, resulting in a negligible variation in the current flowing through the circuit; when reading the superconducting ("1") state, a pulse can be observed at the output due to $R_{NbTiN} = 0$.

The memory could retain the states for over $10^5$ s (see Supplementary Figure S7). Speeds of up to 80 MHz were reached, which is a limit set by our instruments and does not represent the upper limit of our device. The Read operation reached sub-nanosecond speeds (Figure 5c), and the Write operation could be completed within ~5 ns (Figure 5b). A BER of $8 \times 10^{-7}$ was measured with more than 5 million consecutive Read-Write-Read-Reset operations. The memory consumes 56.2 nW in the superconducting state (state "0"), and ~20 nW in the resistive state (state "1"). The energy dissipation was around 1 fJ on the heater to write "1" and ~625 aJ in the channel to read "0". While the mentioned values are comparable with the state-of-the-art nTron memory and logic devices[13], our technology additionally offers non-destructive readout and a small area (~$10^{-2}$ μm²), allowing for a high integration density (see Supplementary Table S1 for an overview).

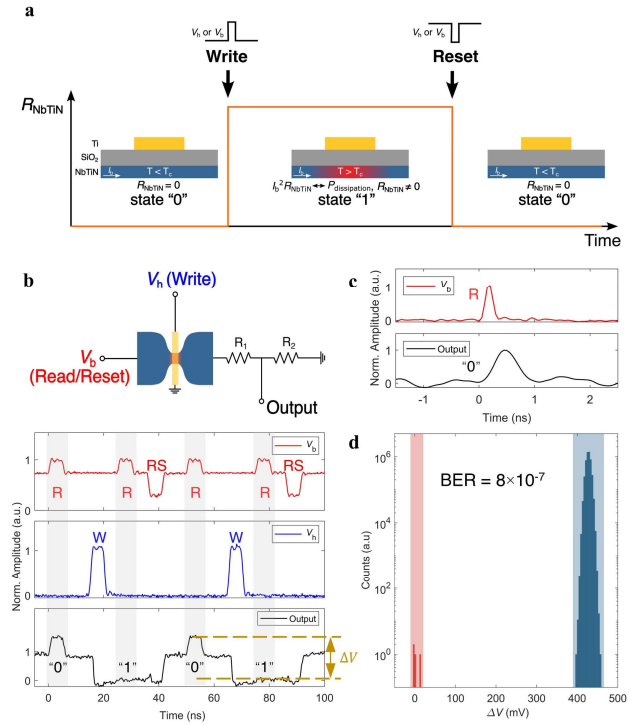

Figure 5: Memory operations of a superconducting thermal switch. **a,** Use of the superconducting thermal switch as a memory cell. **b,** Circuit diagram (top) and experimental characteristics (bottom) of the memory. The output Q is amplified by an AC-coupled amplifier (50 Ω input impedance). $R_1$ = 22 Ω and $R_2$ = 50 Ω are chosen for better impedance matching. Basic memory operations (bottom) are denoted as W for Write, RS for Reset, and R for Read ("0" for reading a superconducting state, where an output pulse is captured, and "1" for reading a resistive state, where no output signal is observed). **c,** Read operation of the state "0" with a sub-nanosecond input pulse. The output pulse has a FWHM of around 550 ps. **d,** Measurement of the BER based on the output voltage difference $\Delta V$ between "0" and "1" state. Incorrect measurements are marked in red and correct measurements in blue.

## 3. Conclusions

In conclusion, we demonstrated a superconducting thermal switch that takes only attojoules to trigger and has a total energy consumption on the order of femtojoules. The multilayered structure is fabrication-friendly and can be scaled down further. Utilizing these superconducting thermal switches, we successfully constructed high-fidelity fundamental logic gates. Compared with standard CMOS logic gates[28], this superconducting logic family requires fewer switching elements and consumes only femtojoules energy per operation. Furthermore, we realized Set-Reset memory operations using a single superconducting thermal switch with a low error rate on the order of $10^{-7}$ and a nanosecond operation speed.

The compactness and reliability of these logic gates is attractive for cryogenic applications, such as, superconducting nanowire single-photon detectors (SNSPDs). SNSPDs are the main choice for optical signal detection in applications such as quantum communication[29], due to their exceptional detection efficiency[30,31] and low time jitter[32], but are difficult to combine into large arrays due to the complexity of their readout circuitry.[33-35] The integration of SNSPDs with the compact, energy-efficient, and fabrication-friendly superconducting electronic components proposed here is a promising route to develop novel readout techniques. Furthermore, the superconducting logic gates and memory cells have the potential to be utilized in high-performance cryogenic digital processors[36-39] as well as hybrid circuits that combine semiconducting and superconducting components[25,40].

**Methods**

**Sample fabrication**

The fabrication of the superconducting thermal switches consisted of 4 steps. First, we created a lift-off pattern with electron-beam lithography for the gold contact pads of the superconducting channel on a wafer with a NbTiN film of 8-10 nm thickness. The thickness of the gold layer was typically 80 nm. Secondly, we etched the NbTiN film to fabricate the superconducting wire with the constriction. The thin NbTiN layer leads to a critical temperature of 8-10 K. Then the whole sample was covered with a silicon dioxide layer of 40~50 nm thickness deposited with ICPCVD (Inductively Coupled Plasma Chemical Vapor Deposition) or PECVD (Plasma Enhanced Chemical Vapor Deposition). This layer primarily serves as an insulating layer but also protects the NbTiN structures from oxidization. Lastly, the Ti heater and its contact pads were patterned in another lift-off process, with a thickness of 80 nm.

**Electrical set-up**

The sample was fixed within a cryostat cooled with liquid helium to achieve a base temperature of around 3 K. The electrical sources and measuring units were connected to the sample externally via coaxial cables, whose characteristic impedance is 50 Ω. It should be noted that the impedance mismatch between the superconducting thermal switches and the coaxial cables can influence the signal transfer efficiency in AC measurements.

The direct current source used in all the electrical measurements consisted of a voltage source and a large bias resistor (usually ~100 kΩ), so that the driving current was approximately constant regardless of the load impedance (usually from 0 to a few kΩ). In the static measurement of the superconducting channel resistance $R_{\mathrm{NbTiN}}$ in Figure 1d, the bias current $I_\mathrm{b}$ and the heater current $I_\mathrm{h}$ were driven with two separate current sources. Then we obtained $R_{\mathrm{NbTiN}}$ by measuring the voltage drop on the NbTiN channel $V_{\mathrm{NbTiN}}$ and using $R_{\mathrm{NbTiN}} = V_{\mathrm{NbTiN}}/I_\mathrm{b}$.

Regarding the AC measurements, the input pulses were generated with an arbitrary waveform generator (Figures 2-4: Tektronix AWG5014C, sampling rate 1.2 G samples/s, Figure 5b: Siglent SDG6032X, sampling rate 1.2 G samples/s) or a short pulse generator (Figure 5c: Highland Technology Model T240). They were attenuated with 20~30 dB and subsequently sent to the sample. The sampling rate of the arbitrary waveform generator limits the rising and the falling time of the input signals to ~800ps. The output signal on the load resistor was coupled to an amplifier through a capacitor with an input impedance of $Z_0 = 50\ \Omega$ and a gain of ~29 dB. All the waveforms were captured with an oscilloscope (Teledyne LeCroy

WaveRunner 640Zi, sampling rate 20 or 40 G samples/s). In order to estimate the energy consumption of the device, the resistance of the heater was measured with a handheld multimeter outside the cryostat, which leads to an overestimation due to the resistance of the bonding pads and the connections between heaters and multimeter probes.

**Data Availability Statement** The data that support the findings of this study are available on request from the authors.

# References

1. Berggren, K. Quantum Computing with Superconductors. *Proceedings of the IEEE* **92**, 1630-1638, doi:10.1109/JPROC.2004.833672 (2004).
2. Wu, Y. *et al.* Strong Quantum Computational Advantage Using a Superconducting Quantum Processor. *Physical Review Letters* **127**, 180501, doi:10.1103/PhysRevLett.127.180501 (2021).
3. Sobolewski, R., Verevkin, A., Gol'tsman, G. N., Lipatov, A. & Wilsher, K. Ultrafast superconducting single-photon optical detectors and their applications. *IEEE Transactions on Applied Superconductivity* **13**, 1151-1157, doi:10.1109/TASC.2003.814178 (2003).
4. Toomey, E. *et al.* Superconducting Nanowire Spiking Element for Neural Networks. *Nano Letters* **20**, 8059-8066, doi:10.1021/acs.nanolett.0c03057 (2020).
5. Khan, S. *et al.* Superconducting optoelectronic single-photon synapses. *Nature electronics* **5**, 650-659, doi:10.1038/s41928-022-00840-9 (2022).
6. Lecocq, F. *et al.* Control and readout of a superconducting qubit using a photonic link. *Nature* **591**, 575-579, doi:10.1038/s41586-021-03268-x (2021).
7. Chen, W., Rylyakov, A. V., Patel, V., Lukens, J. E. & Likharev, K. K. Superconductor digital frequency divider operating up to 750 GHz. *Applied Physics Letters* **73**, 2817-2819, doi:10.1063/1.122600 (1998).
8. Chen, W., Rylyakov, A. V., Patel, V., Lukens, J. E. & Likharev, K. K. Rapid single flux quantum T-flip flop operating up to 770 GHz. *IEEE Transactions on Applied Superconductivity* **9**, 3212-3215, doi:10.1109/77.783712 (1999).
9. Alam, S., Hossain, M. S. & Aziz, A. A cryogenic memory array based on superconducting memristors. *Applied Physics Letters* **119**, 082602, doi:10.1063/5.0060716 (2021).
10. Nagasawa, S., Hinode, K., Satoh, T., Kitagawa, Y. & Hidaka, M. Design of all-dc-powered high-speed single flux quantum random access memory based on a pipeline structure for memory cell arrays. *Superconductor Science and Technology* **19**, S325, doi:10.1088/0953-2048/19/5/S34 (2006).
11. Ryazanov, V. V. *et al.* Magnetic Josephson Junction Technology for Digital and Memory Applications. *Physics Procedia* **36**, 35-41, doi:10.1016/j.phpro.2012.06.126 (2012).

12. McCaughan, A. N. & Berggren, K. K. A Superconducting-Nanowire Three-Terminal Electrothermal Device. *Nano Letters* **14**, 5748-5753, doi:10.1021/nl502629x (2014).
13. Buzzi, A. *et al.* A nanocryotron memory and logic family. *Applied Physics Letters* **122**, 142601, doi:10.1063/5.0144686 (2023).
14. Goldobin, E. *et al.* Memory cell based on a φ Josephson junction. *Applied Physics Letters* **102**, 242602, doi:10.1063/1.4811752 (2013).
15. Suzuki, M., Maezawa, M. & Hirayama, F. Effects of magnetic fields induced by bias currents on operation of RSFQ circuits. *Physica C: Superconductivity* **412-414**, 1576-1579, doi:10.1016/j.physc.2003.12.094 (2004).
16. Tang, G. M., Qu, P. Y., Ye, X. C. & Fan, D. R. Logic Design of a 16-bit Bit-Slice Arithmetic Logic Unit for 32-/64-bit RSFQ Microprocessors. *IEEE Transactions on Applied Superconductivity* **28**, 1-5, doi:10.1109/TASC.2018.2799994 (2018).
17. Katam, N. K. & Pedram, M. Logic Optimization, Complex Cell Design, and Retiming of Single Flux Quantum Circuits. *IEEE Transactions on Applied Superconductivity* **28**, 1-9, doi:10.1109/TASC.2018.2856833 (2018).
18. Razmkhah, S. & Pedram, M. High-density superconductive logic circuits utilizing 0 and π josephson junctions. *Engineering Research Express* **6**, 015307, doi:10.1088/2631-8695/ad27f5 (2024).
19. Bunyk, P., Likharev, K. & Zinoviev, D. RSFQ TECHNOLOGY: PHYSICS AND DEVICES. *International Journal of High Speed Electronics and Systems* **11**, 257-305, doi:10.1142/S012915640100085X (2001).
20. Zhao, Q.-Y. *et al.* A compact superconducting nanowire memory element operated by nanowire cryotrons. *Superconductor Science and Technology* **31**, 035009, doi:10.1088/1361-6668/aaa820 (2018).
21. Butters, B. A. *et al.* A scalable superconducting nanowire memory cell and preliminary array test. *Superconductor Science and Technology* **34**, 035003, doi:10.1088/1361-6668/abd14e (2021).
22. Baghdadi, R. *et al.* Multilayered Heater Nanocryotron: A Superconducting-Nanowire-Based Thermal Switch. *Physical Review Applied* **14**, 054011, doi:10.1103/PhysRevApplied.14.054011 (2020).
23. Chen, B. *et al.* Reconfigurable memlogic long wave infrared sensing with superconductors. *Light: Science & Applications* **13**, 97, doi:10.1038/s41377-024-01424-2 (2024).
24. Nikonov, D. E. & Young, I. A. Benchmarking of Beyond-CMOS Exploratory Devices for Logic Integrated Circuits. *IEEE Journal on Exploratory Solid-State Computational Devices and Circuits* **1**, 3-11, doi:10.1109/JXCDC.2015.2418033 (2015).
25. McCaughan, A. N. *et al.* A superconducting thermal switch with ultrahigh impedance for interfacing superconductors to semiconductors. *Nature electronics* **2**, doi:10.1038/s41928-019-0300-8 (2019).
26. Zheng, K. *et al.* Characterize the switching performance of a superconducting nanowire cryotron for reading superconducting nanowire single photon detectors. *Scientific Reports* **9**, 16345, doi:10.1038/s41598-019-52874-3 (2019).
27. Harrabi, K., Bakare, F., Oktasendra, F. & Maneval, J.-P. Temperature Dependence of the Phonon Escape Time Deduced from the Nucleation Time of Phase Slip Center in Superconducting NbTiN Thin Film. *Journal of Superconductivity and Novel Magnetism* **30**, doi:10.1007/s10948-016-3833-3 (2017).
28. Uyemura, J. P. in *CMOS Logic Circuit Design* (ed John P. Uyemura) 193-258 (Springer US, 2001).

29. You, L. Superconducting nanowire single-photon detectors for quantum information. **9**, 2673-2692, doi:doi:10.1515/nanoph-2020-0186 (2020).
30. Chang, J. *et al.* Detecting telecom single photons with 99.5−2.07+0.5% system detection efficiency and high time resolution. *APL Photonics* **6**, 036114, doi:10.1063/5.0039772 (2021).
31. Colangelo, M. *et al.* Large-Area Superconducting Nanowire Single-Photon Detectors for Operation at Wavelengths up to 7.4 μm. *Nano Letters* **22**, 5667-5673, doi:10.1021/acs.nanolett.1c05012 (2022).
32. Korzh, B. *et al.* Demonstration of sub-3 ps temporal resolution with a superconducting nanowire single-photon detector. *Nature Photonics* **14**, 250-255, doi:10.1038/s41566-020-0589-x (2020).
33. Oripov, B. G. *et al.* A superconducting nanowire single-photon camera with 400,000 pixels. *Nature* **622**, 730-734, doi:10.1038/s41586-023-06550-2 (2023).
34. McCaughan, A. N. *et al.* The thermally coupled imager: A scalable readout architecture for superconducting nanowire single photon detectors. *Applied Physics Letters* **121**, 102602, doi:10.1063/5.0102154 (2022).
35. Hampel, B., Mirin, R. P., Nam, S. W. & Verma, V. B. A 64-pixel mid-infrared single-photon imager based on superconducting nanowire detectors. *Applied Physics Letters* **124**, 042602, doi:10.1063/5.0178931 (2024).
36. Zheng, K. *et al.* A Superconducting Binary Encoder with Multigate Nanowire Cryotrons. *Nano Letters* **20**, 3553-3559, doi:10.1021/acs.nanolett.0c00498 (2020).
37. Lombo, A. E. *et al.* A superconducting nanowire-based architecture for neuromorphic computing. *Neuromorphic Computing and Engineering* **2**, 034011, doi:10.1088/2634-4386/ac86ef (2022).
38. Castellani, M. *et al.* A nanocryotron ripple counter integrated with a superconducting nanowire single-photon detector for megapixel arrays. *arXiv preprint arXiv:2304.11700* (2023).
39. Huang, Y.-H. *et al.* Monolithic integrated superconducting nanowire digital encoder. *Applied Physics Letters* **124**, 192601, doi:10.1063/5.0202827 (2024).
40. Zhao, Q.-Y., McCaughan, A. N., Dane, A. E., Berggren, K. K. & Ortlepp, T. A nanocryotron comparator can connect single-flux-quantum circuits to conventional electronics. *Superconductor Science and Technology* **30**, 044002, doi:10.1088/1361-6668/aa5f33 (2017).

**Contributions**

I. E. Z and H. W conceived and coordinated the project. N. N, I. E. Z and H.W fabricated the samples. S. S, T. D provided the superconducting films. I. E. Z and H. W. designed the experimental setup. H. W. carried out all experiments. All authors contributed to writing the article and reading and approving the final manuscript.

**Acknowledgements**

I. E. Z. and H. W. acknowledge the funding from the FREE project (P19-13) of the TTW-Perspectief research program partially financed by the Dutch Research Council (NWO); I. E. Z. acknowledges funding from the European Union's Horizon Europe research and innovation programme under grant agreement No. 101098717 (RESPITE project) and No.101099291 (fastMOT project).

## Supplementary Information

Supplementary Information is available for this paper.

# Attojoule superconducting thermal logic and memories: Supplementary materials

Hui Wang[1], Niels Noordzij[2], Stephan Steinhauer[3], Thomas Descamps[3], Eitan Oksenberg[2], Val Zwiller[3] & Iman Esmaeil Zadeh[1,2]

[1] Department of Imaging Physics, Delft University of Technology, 2628CN Delft, The Netherlands.
[2] Single Quantum B.V., 2628 CH Delft, The Netherlands.
[3] Department of Applied Physics, Royal Institute of Technology (KTH), SE-106 01 Stockholm, Sweden

## 1. Switching power density

To study the variation in the switching power density of a NbTiN film, several superconducting thermal switches were fabricated on one sample. Each device consisted of a straight superconducting channel with a width $w_{\text{NbTiN}}$, and a metal heater with an area $w_\text{H} \times l_\text{H}$, as schematically displayed in Figure S1a. In the Figure, they are grouped into three categories A, B, and C, according to the superconducting channel width $w_{\text{NbTiN}}$, which was 100 μm, 20 μm, and 5 μm, respectively. A Keithley 2000 multimeter was used to monitor the transition of the superconducting channel using the four-terminal resistance-sensing mode. The influence of the bias current provided by the Keithley on the switching behavior could be neglected since it was around 7 μA, which was much smaller than the critical currents of the superconducting channels. As shown in Figure S1b, devices in the same group switch at a similar switching power density. It can be observed that discrepancies exist among the three groups, which may be attributed to fabrication inaccuracies and variations in the local thermal conduction between the substrate and the cooling stage.

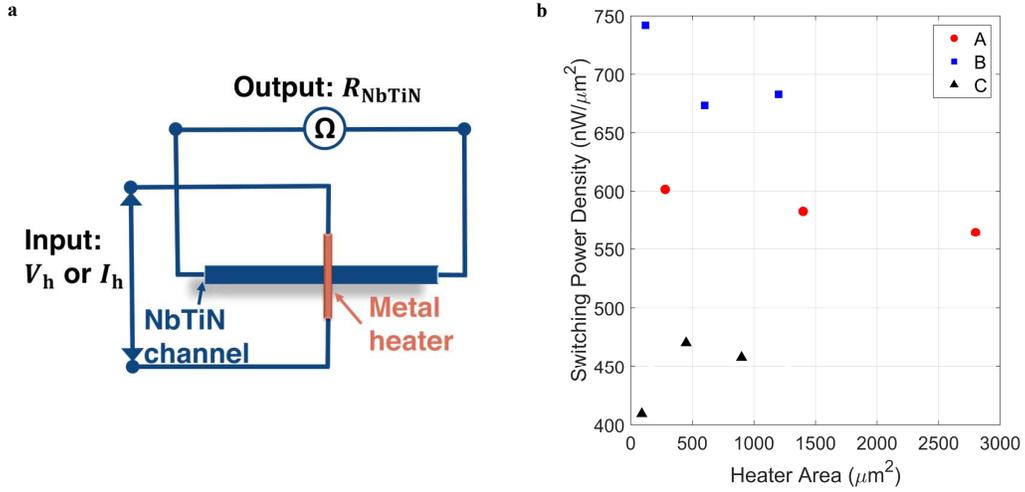

Figure S1: Switching power density measurements. **a,** Measurement set-up to analyze the switching power density of the superconducting thermal switches. **b,** Experimental results for 9 devices on the same sample, categorized into groups A, B, and C according to the superconducting channel width (A: 100 µm, B: 20 µm, and C: 5 µm).

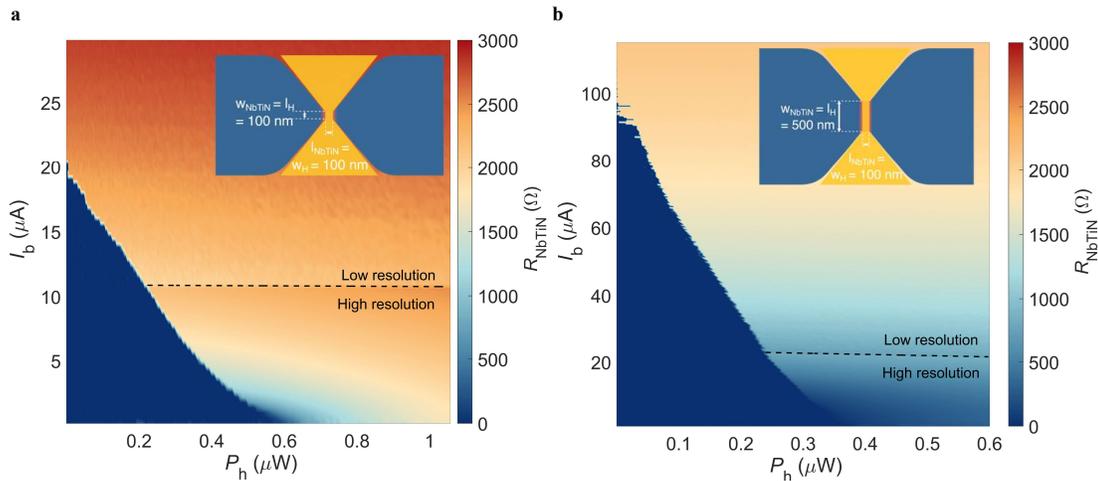

Figure S2: Direct-current characteristics of the superconducting thermal switch. **a** and **b,** Static channel resistance $R_{\text{NbTiN}}$ as a function of the power dissipated by the heater $P_{\text{h}} = I_{\text{h}}^2 R_{\text{h}}$ and the bias current through the channel $I_{\text{b}}$ for two different superconducting thermal switches. The device dimensions are shown in the insets.

## 2. The static and dynamic performances of superconducting thermal switches

Figure S2 shows the static measurements of the channel resistance $R_{\text{NbTiN}}$ of two superconducting thermal switches of different sizes (indicated in the insets of the Figure) fabricated on the same chip. As mentioned in the Methods section, the channel resistance is calculated according to $R_{\text{NbTiN}} = V_{\text{NbTiN}}/I_{\text{b}}$. Given the limited measuring range of the voltmeter, the results are a combination of data acquired with a high-resolution voltmeter at low bias currents and a low-resolution voltmeter at high bias currents. In both figures, the switching heater power at the transition from the superconducting to resistive state decreases as the bias current $I_{\text{b}}$ increases. The switching heater current becomes unstable as $I_{\text{b}}$ approaches the critical current of the NbTiN channel, which is probably due to thermal fluctuations within the set-up. When comparing the switching heater power at a small bias current that is negligible compared to the critical current, it becomes clear that the device in Figure S2a demands a higher switching power than the one in Figure S2b, although the heater size is smaller. However, such a comparison may not be completely meaningful unless the influence of the bonding pads and the connecting wires, which link the critical heater structure to the electrical source, is excluded. To accurately assess the switching power consumption of the superconducting thermal switch, the bonding pads can be made of NbTiN instead of Ti in future experiments. Besides, the difference in switching power could also be attributed to fabrication non-uniformity.

The transient responses of the superconducting thermal switches with various geometries were measured with the same experimental set-up depicted in Figure 2a. Among all the measurements, the lowest dynamic switching energy we achieved was 273.8 aJ (Figure S3a) and the highest switching speed was 200 MHz (Figure S3b).

The device used in Figure S3a has a NbTiN channel of 139 nm (width) × 500 nm (length) and a Ti heater of 100 nm (width) × 3840 nm (length). It was biased with $I_{\text{b}}$ = 31 μA, which was just below its critical current, in order to reach the best dynamic energy consumption of the superconducting switch. According to the input pulse shape (which was attenuated with 20 dB on the heater) and the resistance of the heater $R_h = 594\ \Omega$, the dynamic switching energy measured was $E_h = 273.8$ aJ. The mismatch between the input and the output pulse durations, which might be due to self-heating effects within the NbTiN channel, prevents further investigation into the limit of the switching speed.

Another device tested in Figure S3b has a NbTiN channel of 500 nm (width) × 100 nm (length), and a heater of 100 nm (width) × 500 nm (length). The NbTiN channel has a lower resistance in the normal state, which helps mitigate the self-heating effect and hence increases the switching speed. The experimental results suggest that this device could work at an operation speed of 200 MHz with a pulse duration of 2.5 ns. The Ti heater resistance was $R_h = 282.5\ \Omega$ (100 nm (width) × 500 nm (length)). Thus, the estimated switching energy per operation on the heater was around 580.6 aJ. We believe that the performance of the superconducting thermal switch can be further improved by optimizing the geometries and the electrical circuits.

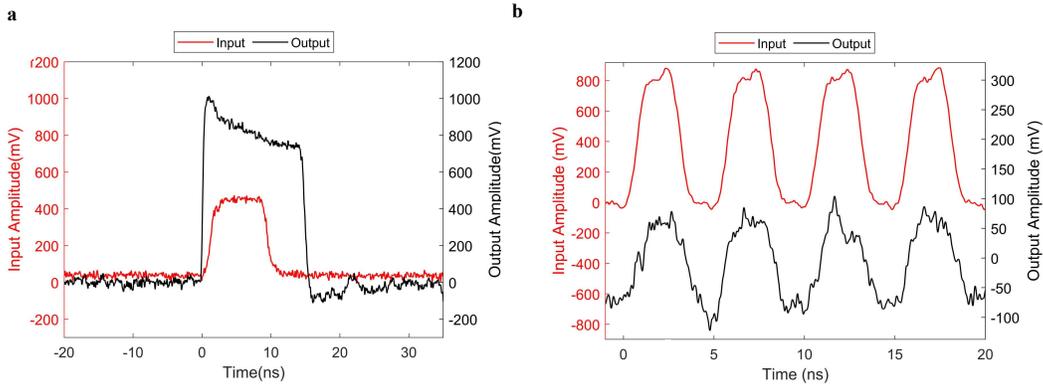

Figure S3: Response of the superconducting switch with the smallest pulse energy and highest switching frequency. **a,** Voltage traces of a superconducting thermal switch with an input pulse energy of 273.8 aJ. The input signal is a 1 MHz square wave with a pulse duration $\tau_{in} \approx 8.33$ ns. **b,** Voltage traces of a superconducting thermal switch operated at 200 MHz. The input signal has a pulse duration $\tau_{in} \approx 2.5$ ns and a duty cycle of 0.5.

## 3. OR and AND gates

The schematics and experimental results of the OR and AND gates are depicted in Figure S4. To build a OR gate, two superconducting thermal switches were connected in series as shown in Figure S4a, and biased with a current source. Here the input impedance of the amplifier $Z_0 = 50\ \Omega$ in parallel with the superconducting switches acted as the load resistor in the experiment, which was coupled through a capacitor. The experimental result in Figure S4b matches the truth table of a OR gate: the output voltage level is LOW when both of the NbTiN channels are superconducting, and becomes HIGH when the superconductivity is broken in either channel because of the joule heating from the heater. Since the normal resistance of the NbTiN channel far exceeds the load resistor, the difference between two HIGH input signals and only one HIGH input signal is not substantial, which is analogous to the observation in the NOR gate. In Figure S4b, the energy dissipation per input pulse on the two heaters was 1.61 fJ and 1.46 fJ. Calculated from the output amplitude and the bias current, the energy consumption in the channel and the load resistor was around 164.9 aJ per operation. Therefore, the whole OR gate required around 3.23 fJ energy dissipation for one operation with both HIGH inputs.

The AND gate is shown in Figure S4c. Here, the two inputs, A and B, activate the NbTiN channel and the heater of the superconducting thermal switch, respectively. A small resistor $R_0 = 22\ \Omega$ was added between the input waveform generator and the NbTiN channel to avoid a short circuit. The load resistor $Z_0$ was placed in parallel with the NbTiN channel via a capacitor. Only when both input levels are HIGH can we obtain a HIGH voltage level at the output, which is consistent with the experimental results in Figure S4c. Assuming both of the inputs were HIGH for the pulse duration $\tau_B = \tau_A \approx 4.2$ ns, the switching energy dissipated in the heater was approximately $\tau_B V_B^2 / R_h = 4.2\text{ ns} \times (6.4\text{ mV})^2 / 194\ \Omega \approx 0.887$ fJ , and the energy consumption in the channel and the biasing circuit was around $\tau_A V_A^2 / (R_0 + Z_0) = 4.2\text{ ns} \times$ (0.28 mV)$^2$/(22 Ω+50 Ω) ≈ 0.00457 fJ. Thus the energy consumption of the whole gate for one output pulse in Figure S4e was around 0.892 fJ.

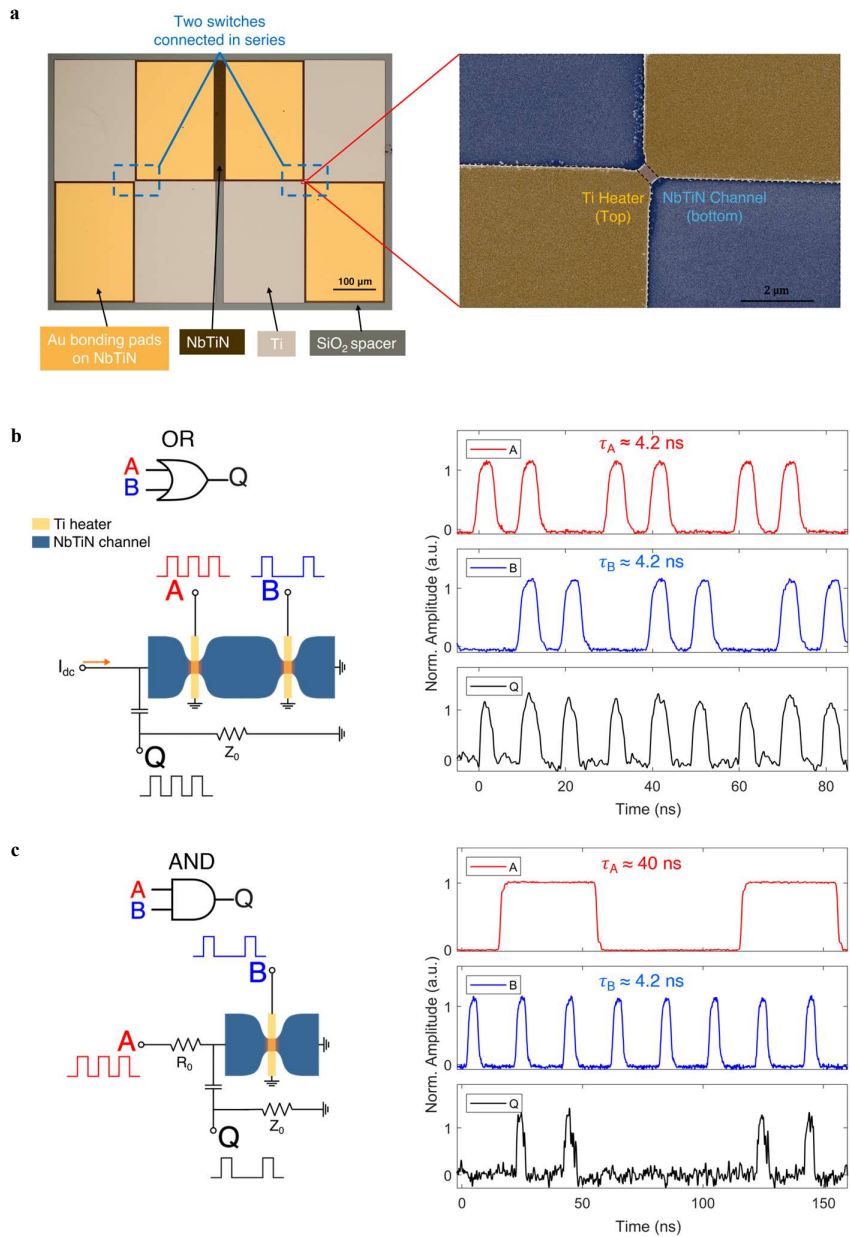

Figure S4: Implementation of the OR and AND gate. **a,** Microscope (left) and SEM (right) image of two superconducting thermal switches whose NbTiN channels are connected in series on the chip. **b-c,** Schematics (left) and rescaled input/output waveforms (right) of the OR (b) and AND (c) gates built using superconducting thermal switches.

## 4. NAND gate implementation

One approach to construct a NAND gate is presented in Figure 3c in the main text, where both the superconducting channel and the heater element carry the input signals. Another feasible method to accomplish a NAND gate is to connect two superconducting switches in parallel, as depicted in Figure S5a. Biasing the two superconducting switches separately is advantageous to ensure a stable switching threshold, as it avoids the current redistribution between two superconducting channels. However, it can be observed from the experimental result in Figure S5b that a gate threshold must be determined so that the low level is only reached when both switches are non-superconducting. Compared with the method shown in Figure 3c, this approach is more complex since it requires more delicate designs of the device geometries and the electrical parameters in multi-stage logic circuits.

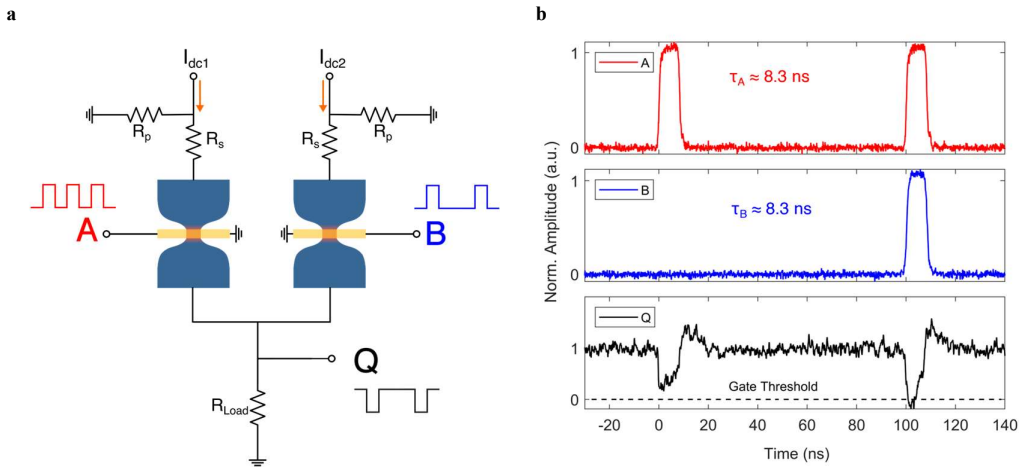

Figure S5: A NAND logic gate composed of two parallel superconducting switches. **a,** Schematic of the circuit. **b,** Rescaled experimental traces of the input and the output signals.

## 5. Memory cell implementations

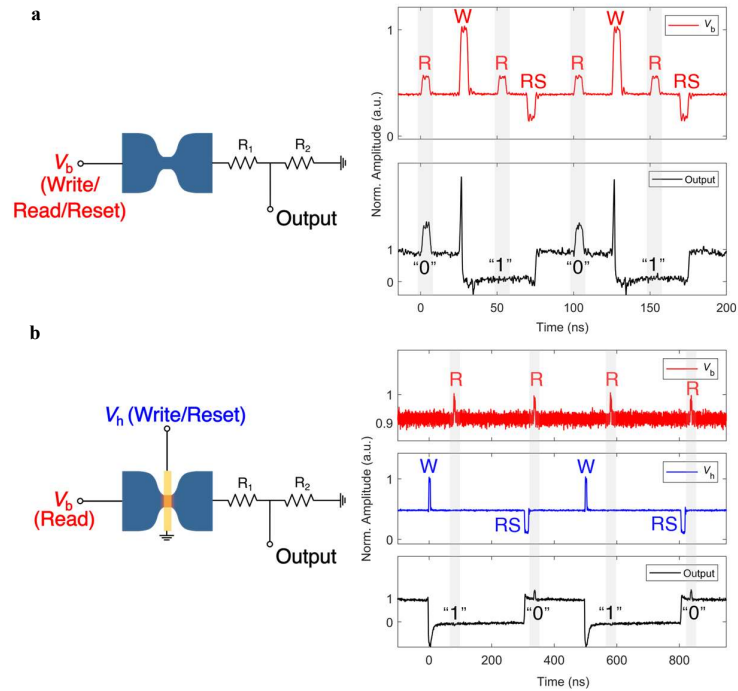

Figure S6: Alternative ways to build a memory cell. Figures on the left depicts the circuit diagrams and figures on the right show the experimental results. **a,** Write (W), Read (R) and Reset (RS) of the memory through the superconducting channel by adjusting the pulse amplitude. **b,** Write (W) and Reset (RS) of the memory via the heater, and Read (R) through the superconducting channel. The output waveform is averaged over 10 measurements for better SNR.

One way to construct a memory cell from superconducting thermal switches is shown in Figure 5, where Write and Reset is done via the heater. There are two other approaches to operate the memory cell, as shown in Figure S6. One approach is to write, read, and reset the memory only through the superconducting channel (Figure S6a). The spike at the writing operation is a consequence of the appearance of the writing pulse and the time delay to switch the channel to its resistive state. The power consumption in the superconducting state dominates in the device, which is around 47 nW. One read operation in superconducting state consumes 572 aJ and one write operation ~1 fJ.
Page **8** of **11**

In the other approach demonstrated in Figure S6b, the memory is written and reset via the heater while reading through the superconducting channel. The total power consumption is approximately 94 nW in state "1" and 112 nW in state "0". A large part of these power consumptions is the static power consumption resulting from the constant bias of the heater. The pulse energies to write and read the device are around 2.82 fJ and 130 aJ, respectively.

**6. Retention time measurement of the memory cell**

In order to monitor the state of the memory cell, we applied Read operations at a frequency of 1 MHz and measured the output amplitudes. Since a pulse can be detected at the output only when the memory is in state "0", the amplitude of the output signal is greater when reading the state "0" than reading the state "1". The measurement taken for over 29 hours is shown in Figure S7, which indicates that both states of our memory cell could be retained for over $10^5$ s.

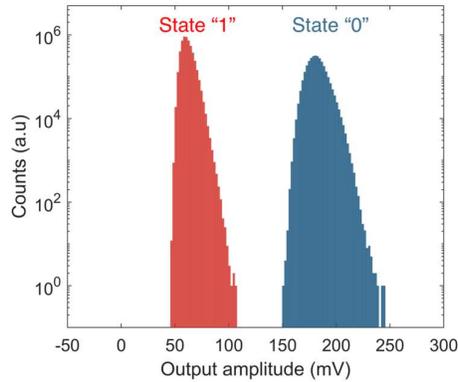

Figure S7: The measurement of the output amplitude of reading the state "0" and the state "1" for over 29 hours.

**7. Comparison with literature values**

In Table S1 we compare the performance of our superconducting thermal logic or memory devices with other superconducting logic or memory devices described in the literatures.

**Supplementary Table S1: Comparison with literature values**

| Ref. | Technology | Type of elements | Power consumption per device | Energy consumption per operation | Speed | Size |
|---|---|---|---|---|---|---|
| **This work** | Superconducting thermal switches | Logic and memory (non-destructive readout) | / | 0.1 to 20 fJ | 100 MHz | $10^{-2}$ µm$^2$ |
| Chen, et al. (2024)[1] | Superconducting nanowire sensors | Memlogic (non-destructive readout) | ~600 µW | / | ~kHz | $10^4$ µm$^2$ |
| Buzzi, et al. (2023)[2] | Superconducting loops with nanocryotrons (nTrons) | Memlogic (destructive readout) | / | 1 fJ | 50 MHz | Few µm$^2$ |
| Butters, et al. (2021)[3] | Superconducting loops with hTrons | Memory (destructive readout) | / | / | Tens of MHz | Few µm$^2$ |
| Zhao, et al. (2018)[4] | Nanowire cryotrons (hTrons and yTrons) | Memory (destructive or non-destructive readout) | / | 10 fJ | Tens of MHz | 21 µm$^2$ |
| Alam, et al. (2021)[5] | Superconducting memristors | Memory | / | 0.1-1 aJ | / | 1-10 µm$^2$ |
| Nair, et al. (2019)[6] | Coupled Superconductor-Insulator-Superconductor Josephson junction arrays | Memory | / | 0.011-0.23 aJ | >10 GHz | / |
| Chen, et al. (1999)[7] | Rapid single-flux-quantum (RSFQ) technology | Logic | / | / | 770 GHz | 0.1 µm$^2$/junction |

**References**

1. Chen, B. *et al.* Reconfigurable memlogic long wave infrared sensing with superconductors. *Light: Science & Applications* **13**, 97, doi:10.1038/s41377-024-01424-2 (2024).
2. Buzzi, A. *et al.* A nanocryotron memory and logic family. *Applied Physics Letters* **122**, 142601, doi:10.1063/5.0144686 (2023).
3. Butters, B. A. *et al.* A scalable superconducting nanowire memory cell and preliminary array test. *Superconductor Science and Technology* **34**, 035003, doi:10.1088/1361-6668/abd14e (2021).
4. Zhao, Q.-Y. *et al.* A compact superconducting nanowire memory element operated by nanowire cryotrons. *Superconductor Science and Technology* **31**, 035009, doi:10.1088/1361-6668/aaa820 (2018).

5. Alam, S., Hossain, M. S. & Aziz, A. A cryogenic memory array based on superconducting memristors. *Applied Physics Letters* **119**, 082602, doi:10.1063/5.0060716 (2021).
6. Nair, N., Jafari-Salim, A., D'Addario, A., Imam, N. & Braiman, Y. Experimental demonstration of a Josephson cryogenic memory cell based on coupled Josephson junction arrays. *Superconductor Science and Technology* **32**, 115012, doi:10.1088/1361-6668/ab416a (2019).
7. Chen, W., Rylyakov, A. V., Patel, V., Lukens, J. E. & Likharev, K. K. Rapid single flux quantum T-flip flop operating up to 770 GHz. *IEEE Transactions on Applied Superconductivity* **9**, 3212-3215, doi:10.1109/77.783712 (1999).